\documentstyle[psfig]{caosp}
\begin{document}
\hauthor{Georges Alecian}
\title{Time-dependent diffusion in A-stars}
\author{Georges Alecian}
\institute{Laboratoire d'Astrophysique Extragalactique et de 
Cosmologie\break 
CNRS (URA173) - Observatoire de Paris - Universit\'e Paris 7\break
DAEC, Observatoire de Meudon
F-92195 Meudon Cedex, France}

\date{December 31, 1997}
\maketitle
\begin{abstract}
Each time diffusion of elements is invoked in explaining abundance anomalies
in a star, this supposes implicitly that a stratification process is in
progress somewhere in that star. This means also, that the element
abundances can still be evolving according to the star's age and fundamental
parameters. Moreover, it has been shown that the superficial abundances may
have complex temporal behavior. This should be detectable through new
observations. In some cases, it may be already apparent in available 
data.

The building up of the elements' stratification is a very difficult
process to study. This is due to the existence of strong non-linearities in the
time-dependent equations, which must be solved numerically. We will discuss
some works that tackle this problem for Ap and Am stars, and we will present
some results concerning mostly Am stars. Future desirable improvements in
these studies will be considered.
\end{abstract}
\section{Diffusion in CP stars}
\subsection{The observational facts}
Here, we shall consider mainly the three well-known groups of Chemically
Peculiar stars on the main-sequence of the HR diagram: FmAm, HgMn
(non-magnetic Ap or weak magnetic field), and magnetic Ap stars. They show
clearly abundance anomalies of metals with respect to solar abundances.
These anomalies extend from factors of about 2 to 10 (in Am stars) and up to
some 10$^{6}$ (in Ap stars).

Many abundance determinations have been done for a large number of Ap and Bp
stars, covering a large sample of effective temperatures, for a large number
of heavy elements. One of the most striking aspects of the observational
results is the rather wide scatter of abundances determined in these stars
for a given element (see for instance Takada-Hidai 1990). This scatter is
partly due to errors in the abundance determinations and to inaccuracies in
the effective temperatures determinations, but not only. Despite this
scatter, a clear correlation with respect to the effective temperature is
well established for some elements like Mn (see for instance Smith and
Dworetsky, 1993).

\subsection{The diffusion model}
It is generally accepted that diffusion processes of elements provide the
best explanation for CP stars anomalies. Of course, microscopic diffusion
interacts with other processes in stars, such as large scale motions
(convection, turbulence, wind) and it cannot be considered alone: the
anomalies are the result of a complex stratification process involving all
these motions.

The diffusion model was first proposed by G. Michaud (1970), it is based on
two basic findings: on the one hand, the abundance anomalies are related to
atmospheric parameters like effective temperature and gravity, on another
hand, there is a clear correlation between the superficial abundance
anomalies and the radiative accelerations (one of the main terms involved in
the diffusion velocity) in stellar external layers. The radiative
accelerations are different according to the elements and they depend on
atmospheric parameters. The anomalies which are found on main-sequence stars
are not found among evolved stars, and this suggests that they are confined 
mostly to external layers. Moreover, in the HR diagram, the CP stars are located where the stars' outer convection zone is supposed to be the
smallest. All of these arguments (with some others which are not discussed
here) give a consistent outline for the diffusion model: when the stars
evolve, the external layers are mixed with deeper ones and the abundance
stratifications are lost.

In the diffusion model, the common property of the CP's is the weakness of
the superficial helium abundance (due to helium gravitational settling).
This helium underabundance is supposed to lead to the decrease (Am stars) or
the disappearance (HgMn and magnetic Ap stars) of the superficial convection
zone. Then, diffusion can occur in layers where stratification time scales
are much shorter than in normal stars.

\section{Stratification processes}
Diffusion is basically a time-dependent process. To determine what
abundances can be observed at the stellar surface one needs to compute the
stratification process which requires solving of the following continuity
equation (valid for test particles only, not for helium):
\begin{equation}
\partial _{t}n+\nabla _{r}\left( \sum\limits_{i}n_{i}\mathbf{V}\mathit{_{ip}}
+n\mathbf{V}\mathit{_{M}}\right) =0
\end{equation}

Where $n$ is the number density of the considered element (and
$n_{i} $  for ions), $\mathbf{V}\mathit{_{M}} $  is a bulk velocity
(stellar wind for instance) and
$\mathbf{V}\mathit{_{ip}} $  is the velocity of ions with respect to protons.

The continuity equation must be solved numerically because it is strongly
non-linear with respect to $n$ (see the study by Alecian and Grappin, 1984,
Alecian, 1986). The diffusion velocity $V_{ip} $ depends on $n_{i} $ 
through the radiative acceleration $g_{i}^{rad} \left( n_{i} (r,t),r\right) $.

This equation is difficult to solve even in the stellar interior
(optically thick, in this case the radiative acceleration is easier to
compute) partly because the diffusion time scale at the bottom (downward boundary)
is very different from the one at the top. On another hand, for Am stars,
the characteristic time of the stratification process in deep layers (below
the outer convection zones) is often of the same order of magnitude than the
time needed to the star's structure to evolve significantly and then, a
detailed computation should take into account the changes of the internal
structure.

\subsection{Optically thick case}

Several works have been done and are in progress to study the stratification
process in A stars. They are based on different approximations.

For Am stars, Alecian (1996) has solved numerically the continuity equation
in the following form (the non-linearities are kept):
\begin{equation}
\partial _{t}C\;+\;\left[ V_{M}-\left\langle D\right\rangle \nabla \ln
\left( N_{p}\left\langle D\right\rangle \right) \right] \nabla
C\;-\left\langle D\right\rangle \nabla ^{2}C\;+\;\frac{1}{N_{p}}\nabla
\left( N_{p}\sum\nolimits_{i}C_{i}v_{ip}\right) =0
\end{equation}

C is the element concentration, $v_{ip}$ is the diffusion velocity of 
ions due to radiative acceleration and gravity.
He has used an approximate formula for the
radiative acceleration which is mainly valid for test particles. On another
hand, he has neglected the effects of star's evolution on the main sequence
and the feedback of element stratification on the stellar model. He has applied 
his
method on calcium diffusion in Am stars (Fig. 1).

\begin{figure}[hbt]
\psfig{figure=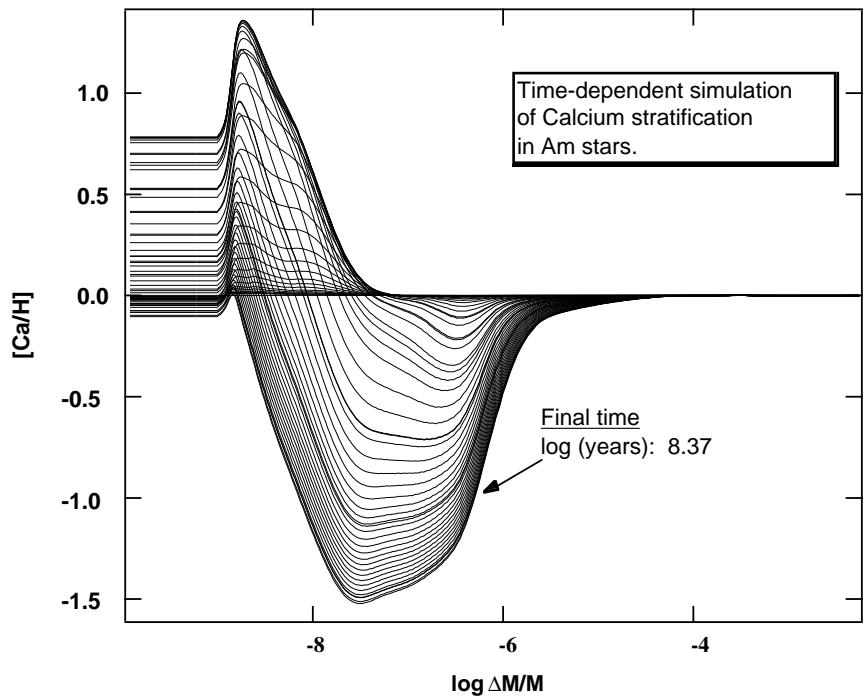,height=9.5cm}
\caption{The logarithm of calcium abundance is plotted versus the logarithm of
the mass fraction above the point of interest. At time t = 0, calcium is
homogeneous and solar, the curves represent the abundance at successive time
steps. The final time corresponds to log (years) = 8.37.}
\label{fp}
\end{figure}

These computations have shown that, according to some values of the mass-loss
rate and thickness of the superficial mixing zone, Ca could be overabundant
in early phases of stratification (see the comparison with observations
by Hui Bon Hoa, this meeting).

Another kind of method has recently been developed by Seaton (1997). He has
proposed a semi-numerical method based on a linearization of the continuity
equation assuming $\mathbf{V}\mathit{_{M}} =0$.
His method has the advantage to use very
accurate radiative acceleration computations but the drawback is that his
method cannot be generalized to any element and to any star.

The group of Montreal (Michaud and collaborators, Turcotte et al, 1998)
has undertaken a very
impressive task of computing the stratification process in a fully consistent
way. The aim is to model the star, its evolution and structure, with
detailed opacities (using OPAL tables). The radiative acceleration is
computed in details for many elements and at each time step, taking into
account the changes of the detailed opacities due to abundance
stratification of all these elements at a time. The continuity equation is solved
numerically. Presently,
as in the previous method, $\mathbf{V}\mathit{_{M}} =0$, but this time
on numerical grounds. This approach is certainly the most promising one,
even if it is very heavy to carry out.

\subsection{Optically thin case}

In the optically thin case (for Ap stars), the situation is much more 
problematic
because radiation transfer, which can no more be solved locally, make the
calculations too huge to be carried out (at the moment): the continuity
equation is coupled to the equation of the radiation transfer. Some attempts
have been done in the past to solve these coupled equations, but they were
rather inaccurate (Alecian and Vauclair 1981, for silicon in Ap stars).

Fortunately, interesting results can be obtained without the need to solve
the continuity equation. This is the case, for example, for manganese by
Alecian and Michaud (1981): the diffusion model allows them to predict the
maximum overabundance that can be observed in HgMn stars versus the
effective temperature. This prediction is based on a quasi ``~zero order~''
approximation (see below), and has been confirmed by the observations of
Smith and Dwortesky (1993).

Generally, the stratification in Ap stars atmospheres is studied with some
strong approximations. For instance, the ``~zero order~'' approximation
consists in solving $g^{rad}\left( n_{equilib.}\right) =g $ which supposes
that the final concentration throughout the medium, is such that the
diffusion velocity is zero everywhere (for instance, Alecian and Artru 1988,
for gallium).
A better approximation is to look for a steady state solution $\nabla
_{r}n_{st}\left( \mathbf{V}\mathit{_{D}}+\mathbf{V}\mathit{_{M}}\right) =0$ (Babel, 1992).
One problem is that one cannot be sure that $n_{equilib.}$ and 
$n_{st}$ are solutions of the continuity equation coupled with the 
transfert equation. Another problem is to know what happens at the 
upper boundary (Babel, 1992), above the photosphere.
For instance, are the elements free to 
leave the atmosphere (role of magnetic field, turbulence, etc.)?

Despite these difficulties, we may try to have some guesses about the
behavior of element stratifications in stellar photosphere (see also Alecian
and Grappin, 1984). In Fig 2, we present a mental picture of what kind of
stratification process could occur in the photosphere of an Ap star. Each
window shows the same section of the photosphere at four successive stages
(t1 to t4). The upper part of the atmosphere is denoted by $\tau $=0 (optical
depth), the bottom ($\tau $=1) is the point where the medium becomes
optically thick.

\begin{figure}[hbt]
\psfig{figure=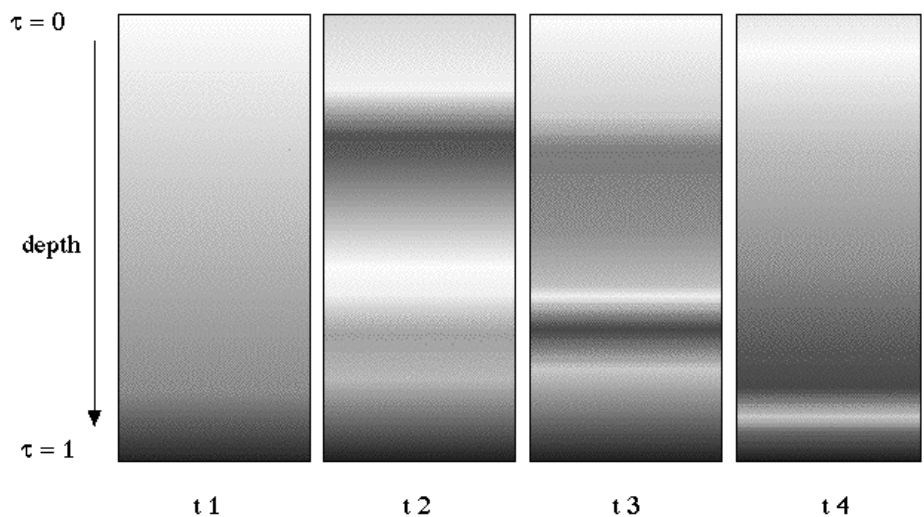,height=7cm}
\caption{This is a mental~picture of what kind of stratification process
could occur in the photosphere of an Ap star (see text). Gray level is
darker when the particle density of the considered element is higher.
}
\end{figure}

Let us consider an element A (different from hydrogen and helium) and assume
that its atomic properties are well fitted to this mental experiment. At
time t1, concentration of A is assumed homogeneous throughout the atmosphere
(the decrease of the gray level represents the decrease of mass density
assuming hydrostatic equilibrium). At time t2, diffusion of A leads to form
a cloud in the upper atmosphere (remember that diffusion time-scale is
shorter when proton density is smaller, then this cloud should appear
first). This cloud is mainly supported by the photons coming from $\tau $=1.
Now, if another cloud of element A forms (at time t3) below the first one,
this second cloud may act as a shield for photons supporting the upper
cloud. Then, the upper cloud is no more supported and falls. This scenario
shows how stratifications built up by diffusion could be unstable in Ap
stars atmospheres. The time scale of such an instability phenomenon 
should be of
the order of the diffusion time-scales in stellar atmospheres, i.e. around
10 to 10$^{3}$ years. No hydrodynamics is involved in this scenario. If this
scenario was confirmed by further theoretical studies (they are presently
in progress), it may contribute in explaining the scatter observed in Ap
stars' abundances.

We have shown that, at the moment, several works have been done to study the
time-dependent diffusion in stellar envelopes. Some of them are rather
accurate but neglect hydrodynamics and are very heavy to carry out. Another
one is less accurate, but takes stellar mass-loss into account, and 
this is more realistic.

Self-consistent modelling, which will determine the evolution on
the main-sequence of an Am star (including all known processes), is within the
scope of the next few years.
However, the studies on the stratification processes in
Ap stars' atmospheres are still embryonic.

In the long run, the goal would be to perform a fully self-consistent modelling
(including hydrodynamics) of the envelope together with the atmosphere. This 
seems to be the only way to answer all the questions raised by observations.


\begin{thebibliography}{}
\bibitem{} Alecian, G.: 1986, {\it Astron. Astrophys.}, 168, 204
\bibitem{} Alecian, G.: 1996, {\it Astron. Astrophys.}, 310, 872
\bibitem{} Alecian, G., Artru, M.C.: 1988, {\it The impact of very high spectroscopy on
stellar physics, IAU symposium N$^\circ$132}, Ed. G.Cayrel de Strobel and
Monique Spite, 235
\bibitem{} Alecian, G., Grappin, R.: 1984, {\it Astron. Astrophys.}, 140, 159
\bibitem{} Alecian, G., Michaud, G.: 1981, {\it Astrophys. J.}, 245, 226
\bibitem{} Alecian, G., Vauclair, S.: 1981, {\it Astron. Astrophys.}, 101, 16
\bibitem{} Babel, J.: 1992, {\it Astron. Astrophys.}, 258, 449
\bibitem{} Michaud, G.: 1970, {\it Astrophys. J.}, 160, 641
\bibitem{} Richer, J., Michaud, G.: 1993, {\it Astrophys. J.}, 416, 312
\bibitem{} Seaton, M.J.: 1997, Mon. Not. R. Astron. Soc., 289, 700
\bibitem{} Smith K.C., Dworetsky M.M. : 1993, {\it Astron. Astrophys.}, 274, 335
\bibitem{} Takada-Hidai, M.: 1990, {\it Evolution of stars, The photospheric abundance
connection}, IAU Symposium N$^\circ$145, Golden Sands, Bulgaria.
\bibitem{} Turcotte, S., Richer, J., Michaud G.: 1998, {\it Astrophys. J.}, in press
\end{thebibliography}
\end{document}